\documentstyle[epsfig,12pt,a4p]{article}

\newcommand{\kkkk}{\mbox{$K^+K^-K^+K^-$ }}
\newcommand{\kstkst}{\mbox{$K^*(892) \overline K^*(892)$ }}
\newcommand{\kpikpi}{\mbox{$K^+K^-\pi^+\pi^-$ }}
\newcommand{\kkpipipi}{\mbox{$K^+K^-\pi^+\pi^-\pi^0$ }}
\newcommand{\pipipipi}{\mbox{$\pi^+\pi^-\pi^+\pi^-$ }}

\begin{document}
\begin{titlepage}
\def\footnoterule{\hrule width 1.0\columnwidth}
\begin{tabbing}
put this on the right hand corner using tabbing so it looks
 and neat and in \= \kill
\> {15 July 1998}
\end{tabbing}
\bigskip
\bigskip
\begin{center}{\Large  {\bf A study of the centrally produced
\kstkst
and $\phi\omega$ systems
in pp interactions at 450 GeV/c}
}\end{center}
\bigskip
\bigskip
\begin{center}{        The WA102 Collaboration
}\end{center}\bigskip
\begin{center}{
D.\thinspace Barberis$^{  4}$,
W.\thinspace Beusch$^{   4}$,
F.G.\thinspace Binon$^{   6}$,
A.M.\thinspace Blick$^{   5}$,
F.E.\thinspace Close$^{  3,4}$,
K.M.\thinspace Danielsen$^{ 11}$,
A.V.\thinspace Dolgopolov$^{  5}$,
S.V.\thinspace Donskov$^{  5}$,
B.C.\thinspace Earl$^{  3}$,
D.\thinspace Evans$^{  3}$,
B.R.\thinspace French$^{  4}$,
T.\thinspace Hino$^{ 12}$,
S.\thinspace Inaba$^{   8}$,
A.V.\thinspace Inyakin$^{  5}$,
T.\thinspace Ishida$^{   8}$,
A.\thinspace Jacholkowski$^{   4}$,
T.\thinspace Jacobsen$^{  11}$,
G.T\thinspace Jones$^{  3}$,
G.V.\thinspace Khaustov$^{  5}$,
T.\thinspace Kinashi$^{  13}$,
J.B.\thinspace Kinson$^{   3}$,
A.\thinspace Kirk$^{   3}$,
W.\thinspace Klempt$^{  4}$,
V.\thinspace Kolosov$^{  5}$,
A.A.\thinspace Kondashov$^{  5}$,
A.A.\thinspace Lednev$^{  5}$,
V.\thinspace Lenti$^{  4}$,
S.\thinspace Maljukov$^{   7}$,
P.\thinspace Martinengo$^{   4}$,
I.\thinspace Minashvili$^{   7}$,
T.\thinspace Nakagawa$^{  12}$,
K.L.\thinspace Norman$^{   3}$,
J.P.\thinspace Peigneux$^{  1}$,
S.A.\thinspace Polovnikov$^{  5}$,
V.A.\thinspace Polyakov$^{  5}$,
V.\thinspace Romanovsky$^{   7}$,
H.\thinspace Rotscheidt$^{   4}$,
V.\thinspace Rumyantsev$^{   7}$,
N.\thinspace Russakovich$^{   7}$,
V.D.\thinspace Samoylenko$^{  5}$,
A.\thinspace Semenov$^{   7}$,
M.\thinspace Sen\'{e}$^{   4}$,
R.\thinspace Sen\'{e}$^{   4}$,
P.M.\thinspace Shagin$^{  5}$,
H.\thinspace Shimizu$^{ 13}$,
A.V.\thinspace Singovsky$^{ 1,5}$,
A.\thinspace Sobol$^{   5}$,
A.\thinspace Solovjev$^{   7}$,
M.\thinspace Stassinaki$^{   2}$,
J.P.\thinspace Stroot$^{  6}$,
V.P.\thinspace Sugonyaev$^{  5}$,
K.\thinspace Takamatsu$^{ 9}$,
G.\thinspace Tchlatchidze$^{   7}$,
T.\thinspace Tsuru$^{   8}$,
M.\thinspace Venables$^{  3}$,
O.\thinspace Villalobos Baillie$^{   3}$,
M.F.\thinspace Votruba$^{   3}$,
Y.\thinspace Yasu$^{   8}$.
}\end{center}

\begin{center}{\bf {{\bf Abstract}}}\end{center}

{
\par
A study of the reactions
$pp \rightarrow p_fp_s(K^+K^-\pi^+\pi^-)$
and $pp \rightarrow p_fp_s(K^+K^-\pi^+\pi^-\pi^0)$
shows evidence for the
\kstkst and $\phi\omega$ channels respectively.
The \kstkst
mass spectrum shows a broad distribution with a maximum near threshold and
an angular analysis shows that it is compatible with having $J^{P}$~=~$2^{+}$.
The behaviour of the cross-section as a function of centre of mass
energy, and the four momentum transfer dependence,
are compatible with what would
be expected if the \kstkst system was produced via double Pomeron exchange.
The $dP_T$ behaviour of the $\phi \omega $ channel
is similar to what has been observed for
all the undisputed
$q \overline q $ states. In contrast, the $dP_T$ behaviour of the
\kstkst final state is similar to what has been observed for the
$\phi \phi$ final state and for previously observed glueball candidates.
}
\bigskip
\bigskip
\bigskip
\bigskip\begin{center}{{Submitted to Physics Letters}}
\end{center}
\bigskip
\bigskip
\begin{tabbing}
aba \=   \kill
$^1$ \> \small
LAPP-IN2P3, Annecy, France. \\
$^2$ \> \small
Athens University, Physics Department, Athens, Greece. \\
$^3$ \> \small
School of Physics and Astronomy, University of Birmingham, Birmingham, U.K. \\
$^4$ \> \small
CERN - European Organization for Nuclear Research, Geneva, Switzerland. \\
$^5$ \> \small
IHEP, Protvino, Russia. \\
$^6$ \> \small
IISN, Belgium. \\
$^7$ \> \small
JINR, Dubna, Russia. \\
$^8$ \> \small
High Energy Accelerator Research Organization (KEK), Tsukuba, Ibaraki 305,
Japan. \\
$^{9}$ \> \small
Faculty of Engineering, Miyazaki University, Miyazaki, Japan. \\
$^{10}$ \> \small
RCNP, Osaka University, Osaka, Japan. \\
$^{11}$ \> \small
Oslo University, Oslo, Norway. \\
$^{12}$ \> \small
Faculty of Science, Tohoku University, Aoba-ku, Sendai 980, Japan. \\
$^{13}$ \> \small
Faculty of Science, Yamagata University, Yamagata 990, Japan. \\
\end{tabbing}
\end{titlepage}
\setcounter{page}{2}
\bigskip
\par
\par
Experiment WA102 is designed to study exclusive final states formed in
the reaction
\begin{equation}
pp \rightarrow p_{f} (X^0) p_{s}
\label{eq:a}
\end{equation}
at 450 GeV/c.
The subscripts $f$ and $s$ indicate the
fastest and slowest particles in the laboratory respectively
and $X^0$ represents the central
system that is presumed to be produced by double exchange processes.
The experiment
has been performed using the CERN Omega Spectrometer,
the layout of which is
described in ref.~\cite{WADPT}.
In previous analyses of other channels it has been observed that
when the centrally produced system has been analysed
as a function of the parameter $dP_T$, which is the difference
in the transverse momentum vectors of the two exchange
particles~\cite{WADPT,closeak},
all the undisputed $q \overline q$ states are suppressed at small
$dP_T$, in contrast to glueball candidates.
In addition, it has recently been suggested~\cite{fec,jmf}
that this suppression could be due to the fact that
the production mechanism is through the fusion of two vector particles.
It is therefore interesting to make a study of the $dP_T$ dependence of
systems decaying to two vectors.
In a recent publication we have performed a study of the centrally produced
$\phi \phi$ final state~\cite{phiphi},
which was found to act
like the glueball candidates that have been observed.
\par
This paper presents a study of the \kstkst final state formed in the reaction
\begin{equation}
pp \rightarrow p_{f} (K^+K^-\pi^+\pi^-) p_{s}
\label{eq:b}
\end{equation}
and the $\phi \omega$ final state formed in the reaction
\begin{equation}
pp \rightarrow p_{f} (K^+K^-\pi^+\pi^-\pi^0) p_{s}
\label{eq:c}
\end{equation}
and represents a factor ten increase in statistics over
previously published data samples~\cite{oldkstkst}.
\par
Events corresponding to reaction~(\ref{eq:b})
have been isolated from the sample of events having six
outgoing
charged tracks
by first imposing the following cuts on the components of the
missing momentum:
$|$missing~$P_{x}| <  14.0$ GeV/c,
$|$missing~$P_{y}| <  0.12$ GeV/c and
$|$missing~$P_{z}| <  0.08$ GeV/c,
where the x axis is along the beam
direction.
A correlation between
pulse-height and momentum
obtained from a system of
scintillation counters was used to ensure that the slow
particle was a proton.
\par
In order to select the \kpikpi system, information
from the {\v C}erenkov counter was used.
An event was accepted if a positive or negative particle
was identified as a K/p and the other particle with the
same charge was consistent with being a $\pi$.
A modified method of Ehrlich et al.~\cite{EHRLICH} ,
has been used to compute the mass
squared of the two highest momentum central particles assuming the other
two particles to be pions.
The resulting distribution is shown in
fig.~\ref{fi:1}a) where a clear peak can be seen at
the kaon mass squared. This distribution has been fitted with
Gaussians to represent the contributions from the
\pipipipi channel and
the \kpikpi channel.
{}From this fit the number of \kpikpi events is found to
be 13950~$\pm$~420.
A cut on the Ehrlich mass squared of
$0.16 \leq M^2 \leq 0.54$~$GeV^2$
has been used to select a sample of \kpikpi events.
The mass of the $K^-\pi^+$ pair is plotted against the mass of the
$K^+\pi^-$ pair in
fig.~\ref{fi:1}b) and shows a strong \kstkst signal.
\par
The two possible $K\pi$ mass combinations are plotted
in fig.~\ref{fi:1}c).  A clear $K^*(892)$
signal can be seen together with an enhancement at 1.4~GeV.
A fit has been performed to this spectrum
using
two Breit-Wigners
to describe the $K^*(892)$ and the peak at 1.4~GeV,
and a background of the form
$(m -m_{th})^a exp(-bm-cm^2)$
where $m$ is the $K\pi$ mass, $m_{th}$ is the threshold mass and
a,b and c are fit parameters.
The result of the fit
gives $m_1$~=~896~$\pm$~2~MeV
and $\Gamma_1$~=~54~$\pm$~3~MeV, compatible with the PDG values~\cite{PDG96}
for the $K^*(892)$ and
$m_2$~=~1436~$\pm$~8~MeV
and $\Gamma_2$~=~196~$\pm$~45~MeV, compatible with the PDG values
for either the $K^*_0(1430)$ or the
the $K^*_2(1430)$.
By selecting one $K\pi$ mass to lie within a band around the $K^*(892)$ mass
(from 0.73 - 1.06~GeV) and plotting the effective mass of the
other pair, the spectrum of fig.~\ref{fi:1}d) was produced.
A strong $K^*(892)$ signal is observed
confirming the presence of the \kstkst final state.
There is no evidence for $K^*(892)K^*(1430)$ production.
\par
The number of events in nine regions around the \kstkst position in the
$K\pi$ scatter plot is shown in fig.~\ref{fi:2}a).
{}From these
numbers and applying a correction for the tails of the $K^*(892)$
the total number of \kstkst events is found to be 2027~$\pm$~113 events.
In order to compare the production rates for $K^*(892) K\pi$  and \kstkst,
the number of
$K^*(892) K\pi$
events has been estimated. This was done
by subtracting twice the number of \kstkst events from the total
number of $K^*(892)$s observed in fig.~\ref{fi:1}c), and this gives
6281~$\pm$~340
$K^*(892) K\pi$
events.
\par
The geometrical acceptance has been found to be similar for \kstkst
and
$K^*(892) K\pi$
production.
After correcting for unseen decay modes of
the $K^*(892)$, the ratio of cross sections is estimated to be
$\sigma(K^*(892) K^\pm\pi^\mp)/\sigma(K^*(892) \overline K^*(892)$)
=~2.06~$\pm$~0.13.
We can also calculate the
number of \kpikpi events that do not include a $K^*(892)$ using the fit
to fig.~\ref{fi:1}a) which gives 5642~$\pm$~550 \kpikpi events.
Although the \kstkst final state is a major component of the
\kpikpi channel it is not the dominant component. In contrast,
the $\phi \phi$ final state was found to be the dominant component
of the \kkkk channel~\cite{phiphi}.
\par
After correcting for
geometrical acceptances, detector efficiencies,
losses due to cuts,
charged kaon decay and unseen decay modes,
the cross-section for
\kstkst production at $\sqrt s$~=~29.1~GeV in the $x_F$ interval
$|x_F| \leq 0.2$ is $\sigma(K^*(892) \overline K^*(892))$~=~85~$\pm$~10~nb.
This can be compared with the cross-sections found in the same interval
at $\sqrt s$~=~12.7 and 23.8~GeV~\cite{oldkstkst} of
67~$\pm$~16~nb and 70~$\pm$~14~nb respectively.
This effectively constant cross-section as a function of energy
is consistent with the \kstkst system being produced by a
Double Pomeron Exchange (DPE) mechanism which is similar to what
was found for the $\phi \phi$ final state~\cite{phiphi}.
\par
In the \kpikpi channel we have also searched for associated
$\phi \rho$ production but have found no evidence for this
channel and can place an upper limit on the cross section of
$\sigma(\phi \rho)$~$<$~0.7~nb
(95~\% confidence limit).
\par
In order to obtain a \kstkst mass spectrum free from background
we have used the nine bin method described above to
calculate the number of \kstkst events in 50~MeV mass intervals.
The resulting \kstkst
effective mass spectrum is shown in fig.~\ref{fi:2}b) and as can be seen
shows a broad distribution with a maximum near threshold.
\par
The angular distributions of the \kstkst system can be used to determine
the spin-parity of the intermediate \kstkst state using
a method formulated by Chang and Nelson~\cite{CHANG}
and Trueman~\cite{TRUEMAN}.
Three angles have to be considered: the azimuthal angle $\chi$, between
the two $K^*(892)$ decay planes and the two polar angles $\theta_1$ and
$\theta_2$
of the $K$s in their respective $K^*(892)$ rest frames
relative to the $K^*(892)$ momenta in the \kstkst rest frame.
\par
For a \kstkst sample of unique spin-parity and free of
background the distribution of $\chi$ takes the form
$dN/d \chi = 1+ \beta cos(2\chi)$
where $\beta$ is a constant which depends only on the spin-parity
of the \kstkst system and is independent of its polarisation.
Similarly
$dN/dcos \theta = 1 + (\zeta/2)(3cos^2\theta - 1)$.
Values of $\beta$ and $\zeta$ for different spin-parity states are
given in table~\ref{ta:1}~\cite{EIGEN}.
The $\chi$ and $cos\theta$ distributions have been obtained
by calculating the number of \kstkst events from the nine bins
in 9 intervals of $\chi$ and 10 intervals of $cos \theta$.
The resulting distributions are shown in
fig.~\ref{fi:3}a) and b)  for $\chi$ and $cos \theta$ respectively.
A chi-squared fit has been performed to these spectra using the
distributions expected for a single
state with a given value of $J^{P}$.
The results of the fit are given in table~\ref{ta:1} and as can be seen
the lowest chi-squared is for a fit using
$J^P$~=~$2^+$ (L=0, S=2) with $\beta$=1/15 and $\zeta$=0.
As can be seen the $J^{P}$~=~$0^{+}$ and $0^{-}$
hypotheses can be clearly ruled out.
A free fit to the distributions has been performed
and gives $\beta$~=~0.17~$\pm$~0.08 and $\zeta$~=~-0.13~$\pm$~0.09.
The spin analysis has been repeated in six slices in mass of the \kstkst
system. The results found are similar to that for the total sample.
The $\phi \phi$ final state
was also found to have $J^P$~=~$2^+$~\cite{phiphi}.
\par
A study of the \kstkst system as a function of the parameter
$dP_T$, which is the difference in the transverse momentum vectors
of the two exchanged
particles~\cite{WADPT,closeak}, has been performed.
The fraction of \kstkst
production has been calculated for
$dP_T$$\leq$0.2 GeV, 0.2$\leq$$dP_T$$\leq$0.5 GeV and $dP_T$$\geq$0.5 GeV and
gives
0.23~$\pm$~0.03, 0.54~$\pm$~0.03 and 0.23~$\pm$~0.02 respectively.
This results in a ratio of production at small $dP_T$ to large $dP_T$ of
1.0~$\pm$~0.16.
This ratio is much higher than that observed~\cite{memoriam} for
the undisputed $q \overline q$ states, which typically have a value for
this ratio
of 0.1, but is similar to that found for glueball candidates
and the $\phi \phi$ final state.
\par
Fig.~\ref{fi:3}c) shows the four momentum transfer at
one of the proton vertices
for the
\kstkst
\newline
system, obtained by calculating the number
of \kstkst events using the nine bin method in six intervals of $t$.
The distribution has been fitted with a single exponential
of the form $exp(-b |t|)$ and yields $b$~=~6.4~$\pm$~0.5~GeV
which is consistent
with what is expected from DPE~\cite{dpet}.
The acceptance corrected azimuthal angle ($\phi$) between the $p_T$
vectors of the two protons
is shown in
fig.~\ref{fi:3}d).
This distribution is similar to that observed
for the $\phi \phi$ final state~\cite{phiphi}, but differs
significantly from that observed
for the $\pi^0$, $\eta$, $\eta^\prime$~\cite{0mpap} and
$\omega$~\cite{3pipap}.
\par
Reaction~(\ref{eq:c})
has been isolated in a similar way to reaction~(\ref{eq:b}) with
the additional requirement of
two $\gamma$s 
reconstructed in the electromagnetic
calorimeter\footnote{The showers associated with the impact of
the charged tracks on the calorimeter
have been removed from the event before the
requirement of only two $\gamma$s was made.}.
The mass of the $K^+K^-$ pair is plotted against the mass of the
$\pi^+\pi^-\pi^0$ system in
fig.~\ref{fi:4}a) and shows evidence for $\phi\omega$ production.
Fig.~\ref{fi:4}b) shows the $K^+K^-$ combination when
the $\pi^+\pi^-\pi^0$ mass combination is required to be in the $\omega$
region
(0.735~$<$~m($\pi^+\pi^-\pi^0$)~$<$~0.835~GeV)
and
fig.~\ref{fi:4}c) shows the $\pi^+\pi^-\pi^0$ combination when
the $K^+K^-$ combination is required to be in
the $\phi$ region (1.01~$<$~m($K^+K^-$)~$<$~1.03~GeV).
The resulting $\phi \omega$ mass distribution is shown in fig.~\ref{fi:4}d).
We have again used the nine bin method to determine the
number of associated $\phi \omega $ events to be 42~$\pm$~14.
After correcting for
geometrical acceptances, detector efficiencies,
losses due to cuts,
charged kaon decay and unseen decay modes,
the cross-section for
$\phi \omega$ production at $\sqrt s$~=~29.1~GeV in the $x_F$ interval
$|x_F| \leq 0.2$ is $\sigma(\phi \omega)$~=~6~$\pm$~2~nb.
\par
A study of the $\phi \omega$ system as a function of the parameter
$dP_T$
has been performed in the three intervals
$dP_T$$\leq$0.2 GeV, 0.2$\leq$$dP_T$$\leq$0.5 GeV and $dP_T$$\geq$0.5 GeV and
gives
0.00~$\pm$~0.06, 0.48~$\pm$~0.20 and 0.52~$\pm$~0.20 respectively.
This results in a ratio of production at small $dP_T$ to large $dP_T$ of
0.00~$\pm$~0.18 compatible to what has been observed for
undisputed $q \overline q$ states~\cite{memoriam}.
\par
In conclusion, a study of the reactions
$pp \rightarrow p_fp_s(K^+K^-\pi^+\pi^-)$
and $pp \rightarrow p_fp_s(K^+K^-\pi^+\pi^-\pi^0)$
show evidence for the
\kstkst and $\phi\omega$ channels respectively.
The \kstkst
mass spectrum shows a broad distribution with a maximum near threshold and
an angular analysis shows that it is compatible with having $J^{P}$~=~$2^{+}$.
The behaviour of the cross-section as a function of centre of mass
energy, and the four momentum transfer dependence,
are compatible with what would
be expected if the \kstkst system was produced via double Pomeron exchange.
The $dP_T$ behaviour of the $\phi \omega $ channel
is similar to what has been observed for
all the undisputed
$q \overline q $ states. In contrast, the $dP_T$ behaviour of the
\kstkst final state is similar to what has been observed for the
$\phi \phi$ final state and for previously observed glueball candidates.
\par
\begin{center}
{\bf Acknowledgements}
\end{center}
\par
This work is supported, in part, by grants from
the British Particle Physics and Astronomy Research Council,
the British Royal Society,
the Ministry of Education, Science, Sports and Culture of Japan
(grants no. 04044159 and 07044098), the Programme International
de Cooperation Scientifique (grant no. 576)
and
the Russian Foundation for Basic Research
(grants 96-15-96633 and 98-02-22032).

\newpage
\newpage
\begin{table}[h]
\caption{The $\beta$, $\zeta$ and chi-squared values for different spins of the
\kstkst system}
\label{ta:1}
\vspace{1in}
\begin{center}
\begin{tabular}{|c|c|c|c|c|c|} \hline
 & & & & & \\
$J^{P}$ & L & S & $\beta$ & $\zeta$ & chi-squared \\
& & & & &\\ \hline
 & & & & & \\
$0^+$  & 0 &  0 &  2/3 & 0 & 54  \\
$0^+$  & 2 &  2 &  1/3 & 1 & 175 \\
$0^-$  & 1 &  1 &  -1 & -1 & 314 \\
$1^-$  & 1 &  1 &  0 & 1/2 & 68 \\
$1^+$  & 2 &  2 &  0 & 1/2 & 68 \\
$2^+$  & 0 &  2 &  1/15 & 0 & 18 \\
$2^+$  & 2 &  0 &  2/3 & 0 & 54 \\
$2^+$  & 2 &  2 &  2/21 & 3/14 & 30 \\
$2^-$  & 1 &  1 &  -2/5 & -1/10 & 64 \\
$2^-$  & 3 &  1 &  -31/5 & -2/3 & 139 \\
 & & & & & \\ \hline
\end{tabular}
\end{center}
\end{table}
\newpage
{ \large \bf Figures \rm}
\begin{figure}[h]
\caption{For the \kpikpi channel a) the Ehrlich mass squared distribution,
b) the lego plot of m($K^-\pi^+$) versus m($K^+\pi^-$),
c) the $K^\pm \pi^\mp$ mass spectrum and
d) the $K^\pm \pi^\mp$ mass spectrum after selecting the
$K^\mp \pi^\pm$ mass to lie in the $K^*(892)$ band.
}
\label{fi:1}
\end{figure}
\begin{figure}[h]
\caption{
For the \kpikpi channel
a) a scatter table of m($K^-\pi^+$) against
m($K^+\pi^-$) in the
\kstkst region and
b) the background subtracted \kstkst mass spectrum.
}
\label{fi:2}
\end{figure}
\begin{figure}[h]
\caption{For the \kstkst channel
a) and b) the $\chi$ and $cos \theta$ distributions
(the superimposed curve represents the $2^+$ (L=0, S=2) wave),
c) the four momentum transfer squared ($|t|$) from one of the proton vertices
and
d) the azimuthal angle ($\phi$) between the two outgoing protons.
}
\label{fi:3}
\end{figure}
\begin{figure}[h]
\caption{For the \kkpipipi channel
a) the scatter plot of m($K^+K^-$) versus m($\pi^+\pi^-\pi^0$),
b) the $K^+ K^-$ mass spectrum after selecting the
$\pi^+\pi^-\pi^0$ mass spectrum to be in the $\omega$ region,
c) the $\pi^+\pi^-\pi^0$ mass spectrum after selecting the
$K^+K^-$ mass spectrum to be in the $\phi$ region and
d) the $\phi \omega$ mass spectrum.
}
\label{fi:4}
\end{figure}
\newpage
\begin{center}
\epsfig{figure=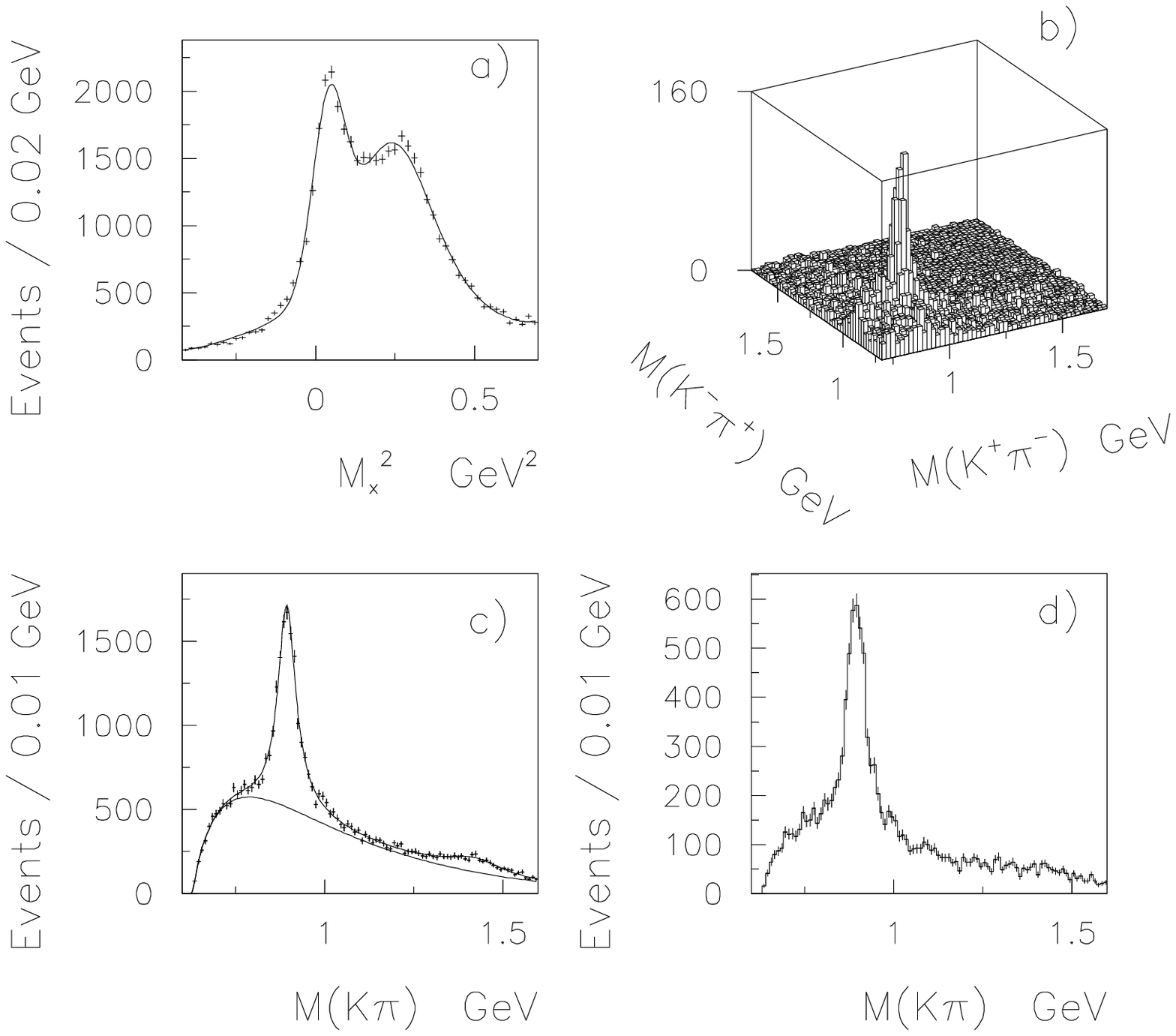,height=22cm,width=17cm}
\end{center}
\begin{center} {Figure 1} \end{center}
\newpage
\begin{center}
\epsfig{figure=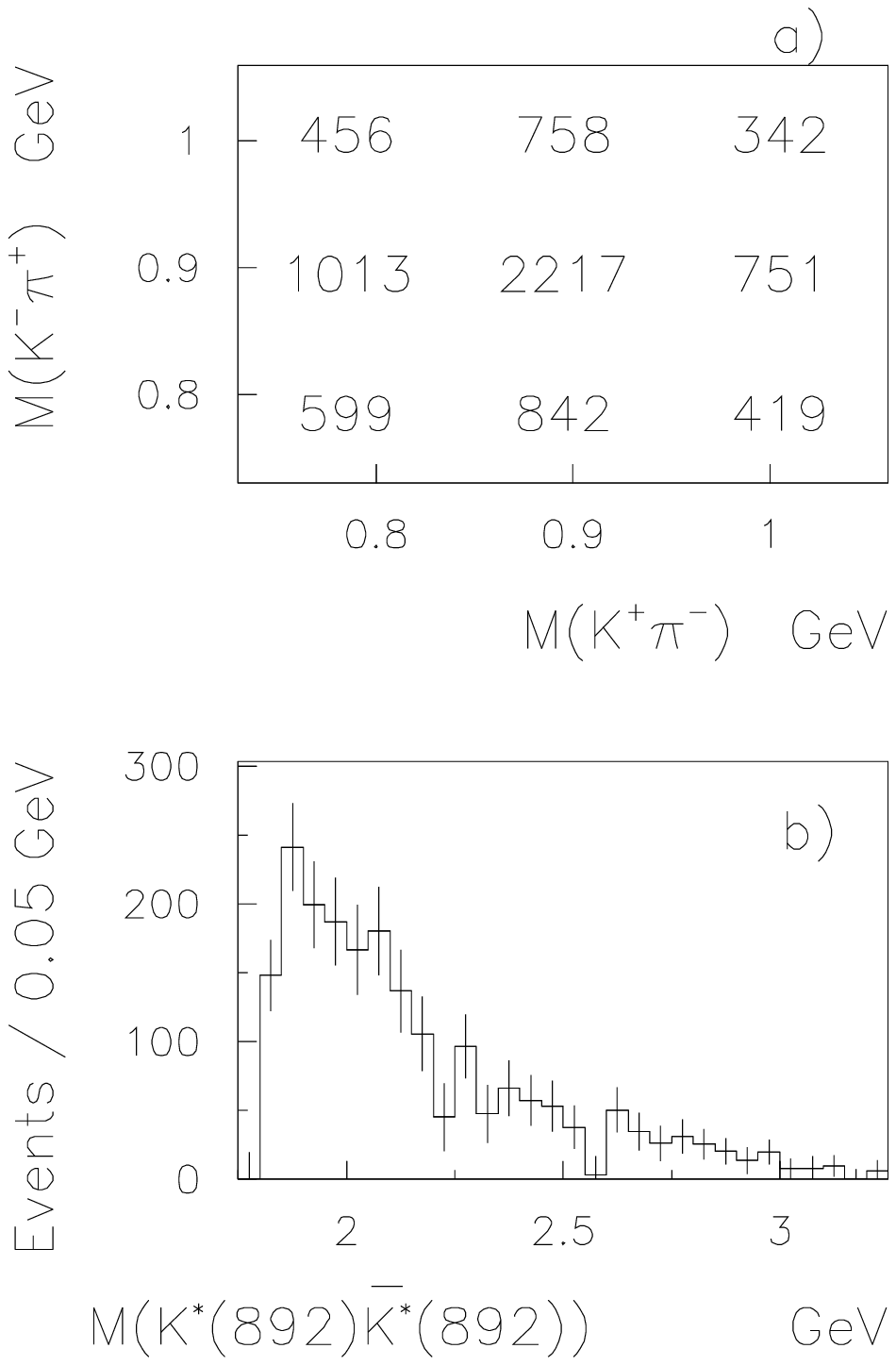,height=22cm,width=17cm}
\end{center}
\begin{center} {Figure 2} \end{center}
\newpage
\begin{center}
\epsfig{figure=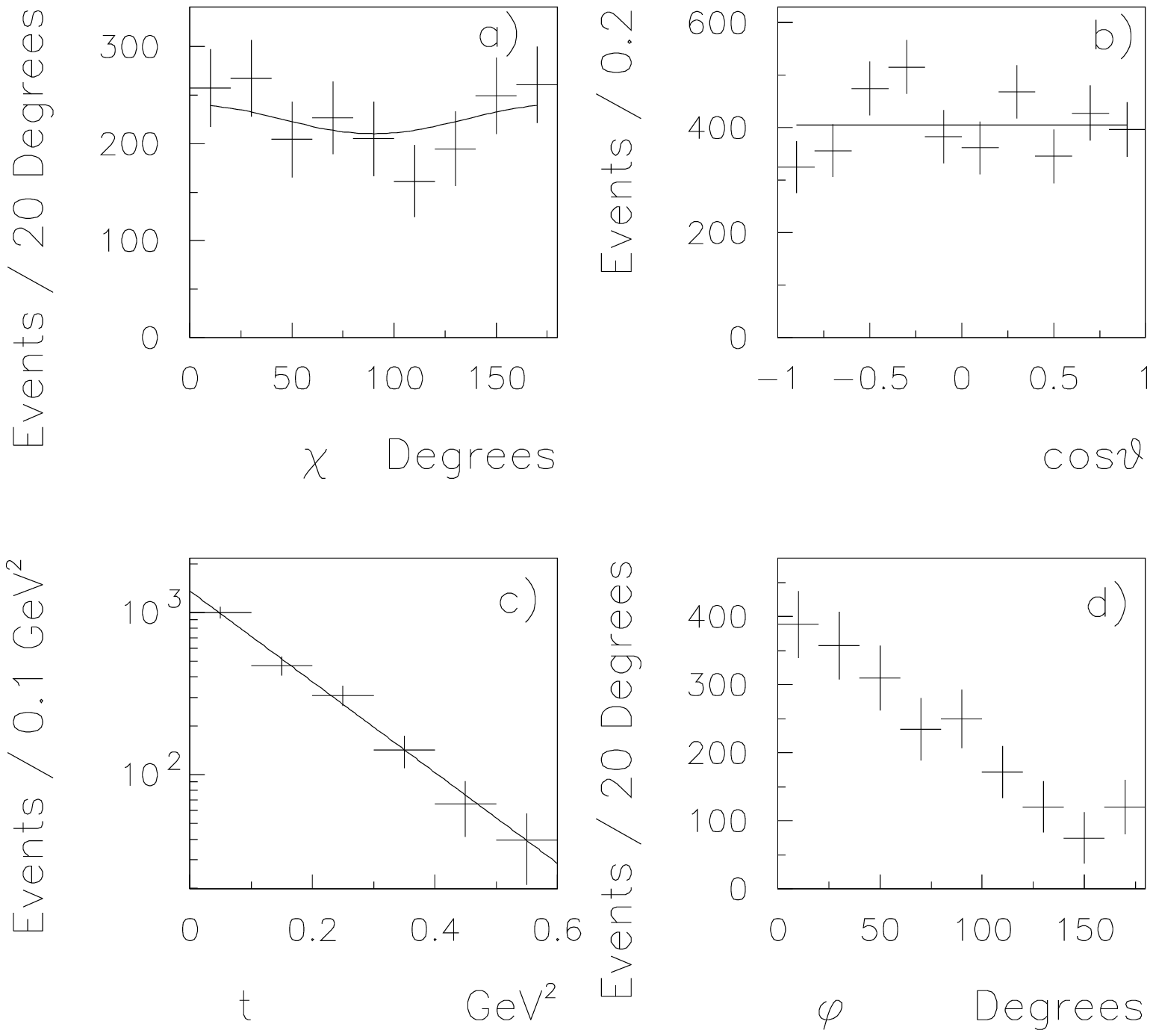,height=22cm,width=17cm}
\end{center}
\begin{center} {Figure 3} \end{center}
\newpage
\begin{center}
\epsfig{figure=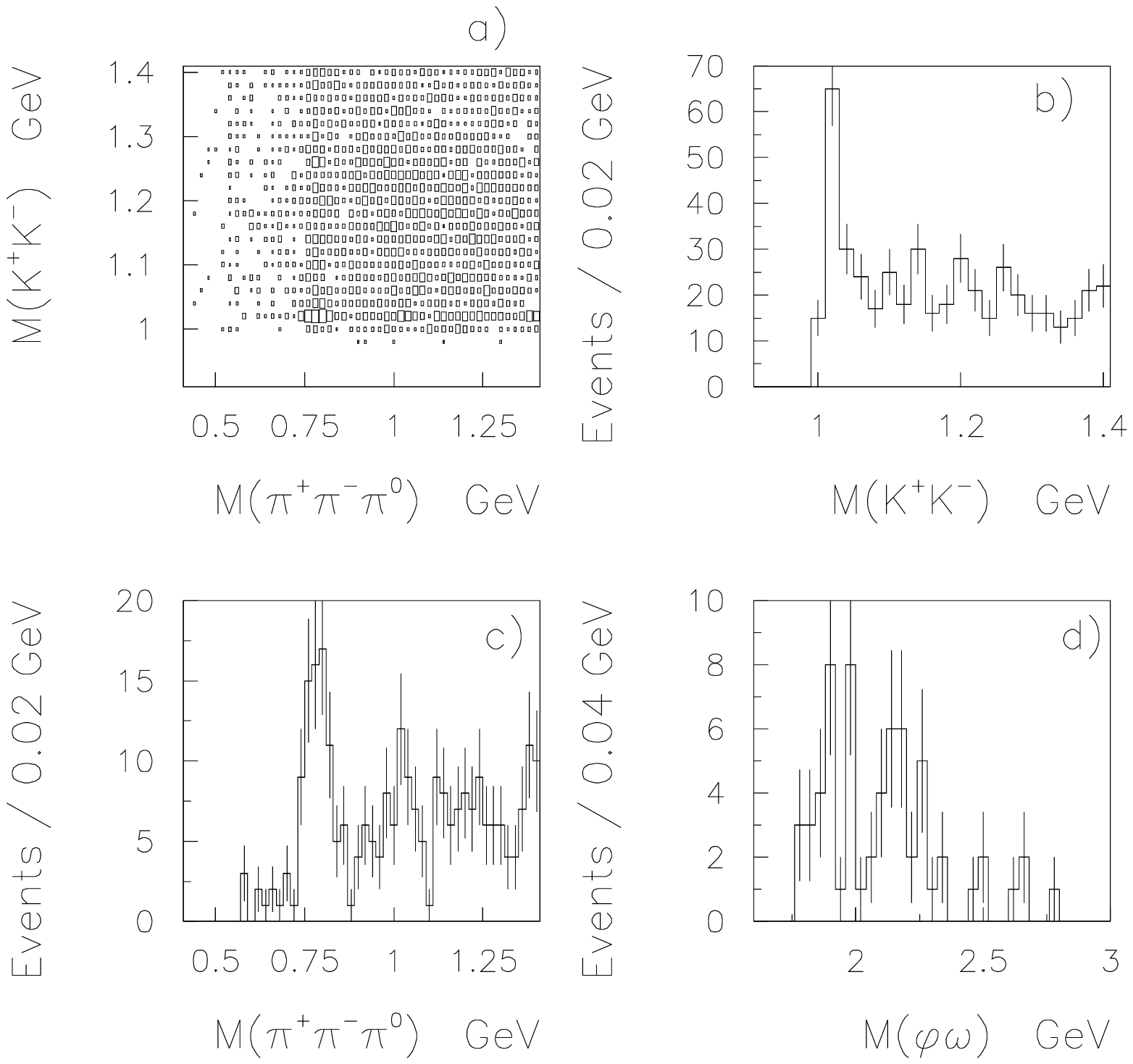,height=22cm,width=17cm}
\end{center}
\begin{center} {Figure 4} \end{center}
\end{document}